\documentclass[reprint,amsmath,amssymb,aps,showpacs,pra]{revtex4-1}
\usepackage{graphicx}
\usepackage{dcolumn}
\usepackage{bm}
\usepackage[usenames,dvipsnames]{color}
\usepackage[utf8]{inputenc}
\usepackage{hyperref}
\graphicspath{{./Files/PDFs/}}

\begin{document}


\title{Study of Dynamo Action in Three Dimensional Magnetohydrodynamic Plasma with Arnold-Beltrami-Childress Flow}

\author{Rupak Mukherjee} 
\email{rupak@ipr.res.in; rupakmukherjee01@gmail.com}
\author {Rajaraman Ganesh}
\email{ganesh@ipr.res.in}
\affiliation{Institute for Plasma Research, HBNI, Bhat, Gandhinagar - 382428, India}

\begin{abstract}
For a three dimensional magnetohydrodynamic (MHD) plasma the dynamo action with ABC flow as initial condition has been studied. The study delineates crucial parameter that gives a transition from coherent nonlinear oscillation to dynamo. Further, for both kinematic and dynamic models at magnetic Prandtl number equal to unity the dynamo action is studied for driven ABC flows. The magnetic resistivity has been chosen at a value where the fast dynamo occurs and the growth rate shows no further variation with the change of magnetic Reynold's number. The exponent of growth of magnetic energy increases, indicating a faster dynamo, if a higher wave number is excited compared to the one with a lower wave number. The result has been found to hold good for both kinematic and externally forced dynamic dynamos where the backreaction of magnetic field on the velocity field is no more negligible. In case of an externally forced dynamic dynamo, the super Alfvenic flows have been found to excite strong dynamos giving rise to the growth of magnetic energy of seven orders of magnitude. The back-reaction of magnetic field on the velocity field through Lorentz force term has been found to affect the dynamics of the velocity field and in turn the dynamics of magnetic field, leading to a saturation, when the dynamo action is very prominent.
\end{abstract}

\maketitle


\section{Introduction}
One of the most interesting open questions of astrophysics is the birth of magnetic field in the cosmos. There are several theoretical models \cite{zeldovich:1983, moffatt:1978} and some of them are tested in laboratory \cite{gailitis:2001, stieglitz:2001, monchaux:2007, ravelet:2008} also, mimicing some aspects of the astrophysical plasma. Amongst the zoo of theoretical models, E N Parker's \cite{parker:1955} theory of {\it dynamo action} is one widely celebrated model. The large-scale magnetic field generation in the `Sun' or in galaxies are mostly attributed to mean-field-dynamo. On the other hand there are astrophysical evidences of small scale magnetic field generation through turbulent fluctuation dynamo \cite{tzeferacos:2018, moll:2011, lovelace:1994, latif:2013, seta:2015, kumar:2014, st:2018}. The origin of dynamo in three dimensional plasma is still poorly understood and still a matter of debate \cite{brandenburg:2012}. In general, the most of the theoretical models employed in this study use the basic equations of  MagnetoHydroDynamics (MHD). The model governs the dynamics of each `{\it fluid element}' - a collisional enough fundamental block of the medium. However, MHD equations describing the plasma in the continuum limit offer fundamental challenges to the analytical solution of the basic equations \cite{morrison:1998}. Hence it is interesting to ask whether there is any finite dimensional description of the subject exists?\\

The authors have shown that in two spatial dimensions for incompressible flows a finite dimensional approach exists and the analytical results were found to fit well with the numerical results obtained earlier \cite{rupak:2018b}. However, the authors also delineate the regimes where the analytical description does not hold good \cite{rupak:2018b}. In three spatial dimensions, the problem becomes more critical to analyse analytically. The phase space of the system being infinite dimensional, in three dimensions, long time prediction of the chaotic trajectories are extremely challenging. But, it was previously shown by the authors that for some typical chaotic flows in three spatial dimensions, the flow and the magnetic field variables are found to reconstruct back to their initial condition - thereby getting trapped in the phase space of the system \cite{rupak:2018c}. The cause of such {\it recurrence} is believed to be the low dimensional behaviour of the single fluid plasma medium for some typical parameters. Most of the short scales were not excited in the system and thus the continuum was acting like a low degrees of freedom medium.\\

Therefore in is natural to ask, is there any way to excite the short scales in a regulated manner as we continuously move in the parameter scale? Finally the question becomes, what happens when all the scales are excited?\\

In the first part of this paper, we address the above questions. We propose a model distinctly showing a continuous transition to self-consistent dynamo from a non-linear coherent oscillation \cite{rupak:2018b}. Though, an analytical description identifying the exact process is still under development, the direct numerical simulation studies of three dimensional chaotic flows support the conjecture. It is known that, in case of a short-scale dynamo, the magnetic field lines frozen to the plasma flows get first stretched along the chaotic velocity flows - then gets twisted and folded back \cite{ott:1998}. Such processes introduce generation of short scales into the system giving birth of dynamos classified as STF dynamo \cite{vainshtein:1996, vainshtein:1997}. Though it is well known that a small but non-zero resistivity affects the plasma relaxation because of reconnection process \cite{woltjer:1958, taylor:1974, parker:1979, taylor:1986, qin:2012, rupak:2018d}, we choose flows with finite viscosity and resistivity showing the robustness of our results. We see a continuous growth of magnetic energy at the cost of kinetic energy and thereafter decrease of the magnetic energy through reconnection process and converting back to kinetic energy. We move in parameters and find the reconnection to occur with less probability and growth of magnetic energy. Thus we move in parameter space and in one limit observe coherent nonlinear oscillation of kinetic and magnetic energy within the premise of single fluid magnetohydrodynamics and in the other limit observe dynamo action to occur.\\

From our previous study \cite{rupak:2018c}, we choose flows that do not recur and thus even though there is an energy exchange between the kinetic and the magnetic modes, the trapping in phase space does not occur even in the opposite limit of the dynamo.\\

As a test case we choose Arnold-Beltrami-Childress (ABC) flow since it is already known to produce fastest dynamos in the kinematic regimes \cite{alexakis:2011} and is non-recurrent \cite{rupak:2018c}. We reproduce the works of Galloway and Frisch \cite{galloway:1986} in the kinematic regime and further check our results in the self-consistent regime with the work of Sadek {\it et al} \cite{alexakis:2016}. We use the well-benchmarked three dimensional MHD code G-MHD3D \cite{rupak:2018, rupak:2018a} capable of direct numerical simulation of weakly compressible single fluid MHD equations.\\

In the second part of the paper, the growth of magnetic energy (dynamo action) under the action of externally driven ABC flow has been studied. The analysis can be divided into two parts, {\it i)} the linear kinematic regime where the plasma flow stretches the magnetic field lines giving rise to an exponential growth of magnetic field and {\it ii)} the nonlinear regime where the magnetic field generates Lorentz force strong enough to modify the topology of the background plasma flow, resulting in a saturation of the growth rate of the magnetic field \cite{childress:2008, brandenburg:2005}. \\

In the linear kinematic regime, the growth of magnetic field is found to exponentially rise without bound after crossing a critical threshold Reynold's number ($Rm_c$). When the backreaction is not negligible, (we dub this case as {\it dynamic dynamo}) the dynamo saturates when it enters the nonlinear regime. At very low magnetic Prandtl number ($P_m \ll 1$) the dynamic dynamo has been studied in detail \cite{brandenburg:2001, steenbeck:1966, krause:2016, mininni:2007, mininni:2005, childress:2008, galanti:1992, ponty:1995, archontis:2003, frick:2006}. A detailed review of the ABC flow leading to dynamo action is due to Galloway \cite{galloway:2012}. However, in the solar convection zone, unprecedentedly high resolution simultion with Prandtl number unity ($P_m = 1$) has been shown to produce global-scale magnetic field even in the regime of large Reynolds numbers \cite{hotta:2016}. For $P_m = 1$ and Reynold's number $Rm$ $\textgreater$ $Rm_c$, we extend our search for faster dynamos (now dynamic one) when the initial velocity and forcing scales contain higher wave-number. We also address the cases where the Alfven speed and sonic speed differ significantly. We see that the backreaction of the magnetic field alters the ABC flow profile and thereby the growth of magnetic energy itself gets affected. We notice three distinct growth rates in the magnetic energy solely because of the inclusion of the backreaction of the magnetic field on the velocity field and vice-versa.\\

In particular, the following aspects of dynamo action under ABC flow have been studied in detail:
\begin{itemize}
\item We identify crucial parameter that controls the generation of short scales and thereby lead to dynamo action.
\item We find that, above the critical magnetic Raynold's number ($Rm_c$), the growth rate of kinematic dynamo process increases with velocity scale containing higher wave numbers.
\item We also find that, above $Rm_c$, the growth rate of self-consistent or dynamic dynamo also increases, with velocity and forcing scales containing higher wave numbers.
\item For super Alfvenic flows, when strong dynamo action occur, the effect of interplay of energy between magnetic and kinetic modes leads to saturation of the growth of magnetic energy at late times.
\item For ABC flow to start with, both for kinematic as well as dynamic dynamo action, the magnetic energy is primarily contained in the intermediate scales.
\end{itemize}


\section{Governing Equations} 
The single fluid magnetohydrodynamic (MHD) description of a plasma is quite incomplete but has been found to serve aptly to explain many phenomena observed in laboratory and astrophysical systems. Thus under certain criteria, the plasma dynamics is believed to be well modelled through single fluid MHD equations. The two different charge species (electrons and ions) are assumed to form a single fluid because of the negligible mass of the electrons. A fluid element is assumed to be much larger than the length scale of separation between the two different charge species. Also the timescale at which the phenomena are observed are quite longer than the gyrofrequency of each of the charge species. Thus no large scale electric field is produced or sustained in the timescale of interest.\\

The basic equations governing the dynamics of the magnetohydrodynamic fluid are as follows: 
\begin{eqnarray}
&& \label{density} \frac{\partial \rho}{\partial t} + \vec{\nabla} \cdot \left(\rho \vec{u}\right) = 0\\
&& \frac{\partial (\rho \vec{u})}{\partial t} + \vec{\nabla} \cdot \left[ \rho \vec{u} \otimes \vec{u} + \left(P + \frac{B^2}{2}\right){\bf{I}} - \vec{B}\otimes\vec{B} \right]\nonumber \\
&& \label{velocity} ~~~~~~~~~ = \mu \nabla^2 \vec{u} + \rho\vec{f}\\
&& \label{Bfield} \frac{\partial \vec{B}}{\partial t} + \vec{\nabla} \cdot \left( \vec{u} \otimes \vec{B} - \vec{B} \otimes \vec{u}\right) = \eta \nabla^2 \vec{B}\\
&& \text{where} ~~ P = C_s^2 \rho ~ \text{and} ~ \vec{f} = \begin{pmatrix}
A \sin(k_f z) + C \cos (k_f y) \\
B \sin(k_f x) + A \cos (k_f z) \\
C \sin(k_f y) + B \cos (k_f x)
\end{pmatrix}. \nonumber 
\end{eqnarray}
In the above system of equations, $\rho$, $\vec{u}$, $P$ and $\vec{B}$ are the density, velocity, kinetic pressure and the magnetic field of a fluid element respectively. $\mu$ and $\eta$ denote the coefficients of kinematic viscosity and magnetic resistivity. We assume $\mu$ and $\eta$ are constants throughout space and time. The symbol ``$\otimes$'' represents the dyadic between the two vector quantities.\\

The kinetic Reynold's number ($Re$) and magnetic Reynold's number ($Rm$) are defined by $Re = \frac{U_0 L}{\mu}$ and $Rm = \frac{U_0 L}{\eta}$ where $U_0$ is the maximum velocity of the fluid medium to start with and $L$ is the system length.\\

We also define the sound speed of the fluid medium as $C_s = \frac{U_0}{M_s}$, where, $M_s$ is the sonic Mach number of the fluid. We assume it to be uniform throughout the space and time. The Alfven speed is determined from the relation $V_A = \frac{U_0}{M_A}$ where $M_A$ is the Alfven Mach number of the plasma medium. The initial magnetic field present in the plasma is determined from relation $B_0 = V_A \sqrt{\rho_0}$, where, $\rho_0$ is the initial density profile of the fluid.
   

\section{Parameter Details}
\label{sec:parameter}
The first results of Galloway and Frisch \cite{galloway:1984} showed critical dependency on the magnitude of magnetic resistivity ($Rm$). Later this result was further tested and reproduced with much greater accuracy and resolution by in several other independent studies \cite{bouya:2013,bouya:2015}. The observation that, within a $2 \pi$ periodic box, for algorithms depending on spectral solvers, the smallest features in magnetic field are on scales of order $Rm^{-1/2}$ indicates that the grid size required to resolve a given $Rm$ scales like $Rm^{1/2}$ \cite{galloway:1986}. We choose, $N = 64$ which resolves $Rm$ upto $4096$ and keep our parameters fixed at $Rm = 450$ (where the growth rate of dynamo was found to get saturated \cite{galloway:1986}) well within the resolution threshold. Also the result of Sadek {\it et al} \cite{alexakis:2016} confirms that most of the kinetic and magnetic energy content remains within the large scales, even when the driving wave-number is kept at intermediate scales (at least upto $k_f = 16$). This sets limit to our choice of maximum driving wave number ($k_f = 16$) at the grid resolution $N = 64$.

Throughout our simulation, we set $N = 64$, $L = 2 \pi$, $\delta t = 10^{-4}$, $\rho_0 = 1$. For some test runs the grid resolution is increased to $N = 128$ for both kinematic and dynamic cases but we found no significant variation of the physics results. The initial magnitude of density ($\rho_0$) is known to affect the dynamics and growth rate of an instability in a compressible neutral fluid \cite{bayly:1992, terakado:2014}. However, in present case we keep the initial density fixed ($\rho_0 = 1$) for all the runs. We check our code with smaller time stepping ($\delta t$) keeping the grid resolution $N = 64$. No deviation from the results were observed with such test runs. 

The kinematic viscosity is controlled through the parameters $Re$ and to guarantee similar decay of kinetic and magnetic energy, we set $Re = Rm$ everywhere. Next we vary the Alfven speed through $M_A$ and observe the effect of these parameters on the dynamo action. We also change the magnitude of forcing by controlling the values of $A, B, C$ and the length-scale of forcing through $k_f$. 

The OpenMP parallel MHD3D code is run on 20 cores for 9600 CPU hours for a single run with parameters mentioned above and got the following results.


\section{Simulation Results}
\label{sec:simulation_results}

\subsection{Initial Profile of Density, Velocity and Magnetic Field}
We start with the initial condition $\rho = \rho_0$ as a uniform density fluid, the initial velocity profile as $u_x = U_0 [A \sin(k_f z) + C \cos (k_f y)]$, $u_y = U_0 [B \sin(k_f x) + A \cos (k_f z)]$, $u_z = U_0 [C \sin(k_f y) + B \cos (k_f x)]$ and the initial magnetic field as $B_x = B_y = B_z = B_0$. We keep the initial profiles of all the fields identical throughout our paper unless otherwise stated. 


\subsection{Transition to Dynamo}
\begin{figure}
\includegraphics[scale=0.65]{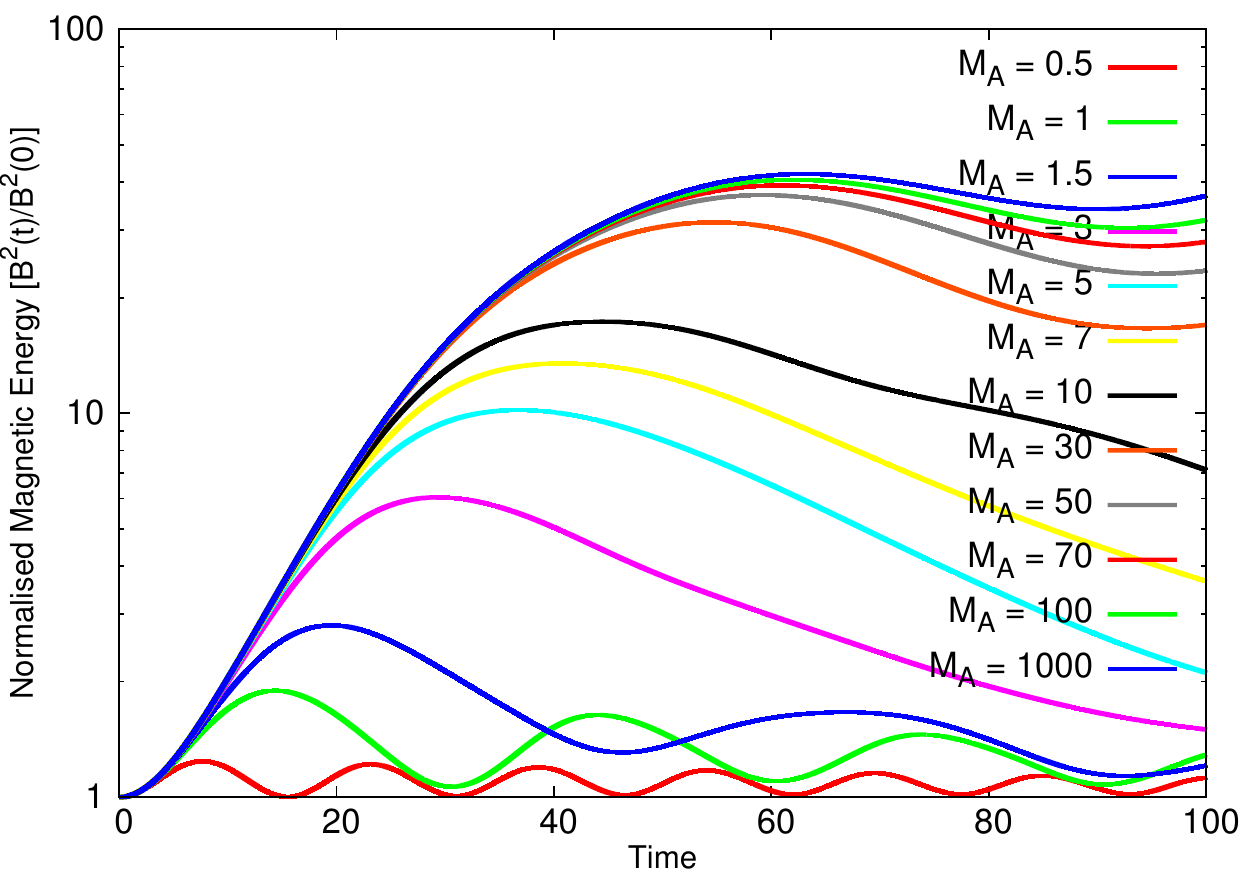}
\caption{(Color online) Transition to dynamo with the increase of $M_A$ from coherent nonlinear oscillation.}
\label{dynamo_MA}
\end{figure}

For $M_A \sim 1$, a coherent nonlinear oscillation is reproduced as reported earlier \cite{rupak:2018b}. As $M_A$ is moved from unity, the oscillation persists alongwith the generation of other modes into the system. Thus as can be found from Fig. \ref{dynamo_MA}, the magnetic energy does not come back to its initial value after one period of oscillation. Upon further increment of Alfven Mach number, the linear dependency of the  frequency of oscillation breaks down and persistent magnetic field starts to generate. Finally, the growth of magnetic energy reaches a maximum. From Fig. \ref{dynamo_MA}, it can be seen that, the normalised magnetic energy at $M_A = 10^2$ \& $10^3$ does not differ significantly, indicating a saturation of the growth. However, such saturation does not occur in the driven cases where, the plasma is driven continuously using an external drive which pumps in kinetic energy to the system. Such phenomena is further explored in the next section of the paper.

\subsection{Kinematic Dynamo}
\label{subsec:kinematic_dynamo}
The phenomenon of magnetic energy growing exponentially with time for a statistically steady flow, where the velocity field is held fixed in time, is called, kinematic dynamo action. Arnold-Beltrami-Childress (ABC) flow being a steady solution of Euler equation, sets the premise to study the kinematic dynamo problem. For ABC flow kinematic dynamo was first obtained by Arnold {\it et al} \cite{arnold:1983} at magnetic Reynold's number ($Rm$) between $9$ and $17.5$. Galloway {\it et al} \cite{galloway:1984} found a more efficient dynamo effect with much higher growth rate after $Rm = 27$ breaking certain symmetries of the flow. Later on, the study had been extended for the parameters where $A$, $B$, $C$ are not equal \cite{galloway:1986}. The threshold $Rm$ for a kinematic dynamo has been well explored \cite{ponty:2005, mininni:2007, schekochihin:2005, iskakov:2007}. The real part of the growth rate of the magnetic energy for increasing $Rm$ is found to increase while the imaginary part decreases continuously \cite{galloway:1986}. ABC flows with differnet forcing scales ($k_f \neq 1$) providing kinematic dynamo has been explored by Galanti {\it et al} \cite{galanti:1992} and more recently by Archontis {\it et al} \cite{archontis:2003}. 

First we reproduce the previous results and then choose an optimal set of working parameters. The motivation behind choosing the parameters are explained in the previous section. We give the following runs (Table \ref{parameter_run_kinematic}) to explore the parameter regime of a kinematic dynamo problem.
\begin{table}[h!]
\centering
\begin{tabular}{ |c|c|c|c| }
 \hline
 Name & $k_f$ & $M_A$ & $Rm$ \\
 \hline
 KF1 & $1$ & $1000$ & $450$ \\
 KF2 & $2$ & $1000$ & $450$ \\ 
 KF3 & $4$ & $1000$ & $450$ \\ 
 KF4 & $8$ & $1000$ & $450$ \\ 
 KF5 & $16$ & $1000$ & $450$ \\
 KM1 & $1$ & $100$ & $450$ \\
 KR1 & $1$ & $10$ & $450$ \\ 
 KR2 & $1$ & $10$ & $200$ \\  
 KR3 & $1$ & $10$ & $120$ \\   
 KM2 & $1$ & $1$ & $450$ \\   
 KM3 & $1$ & $0.1$ & $450$ \\ 
 \hline
\end{tabular}
\caption{Parameter details with which the simulation has been run for kinematic dynamo problem.}
\label{parameter_run_kinematic}
\end{table}


\subsubsection{Effect of Magnetic Resistivity}
\begin{figure}
\includegraphics[scale=0.65]{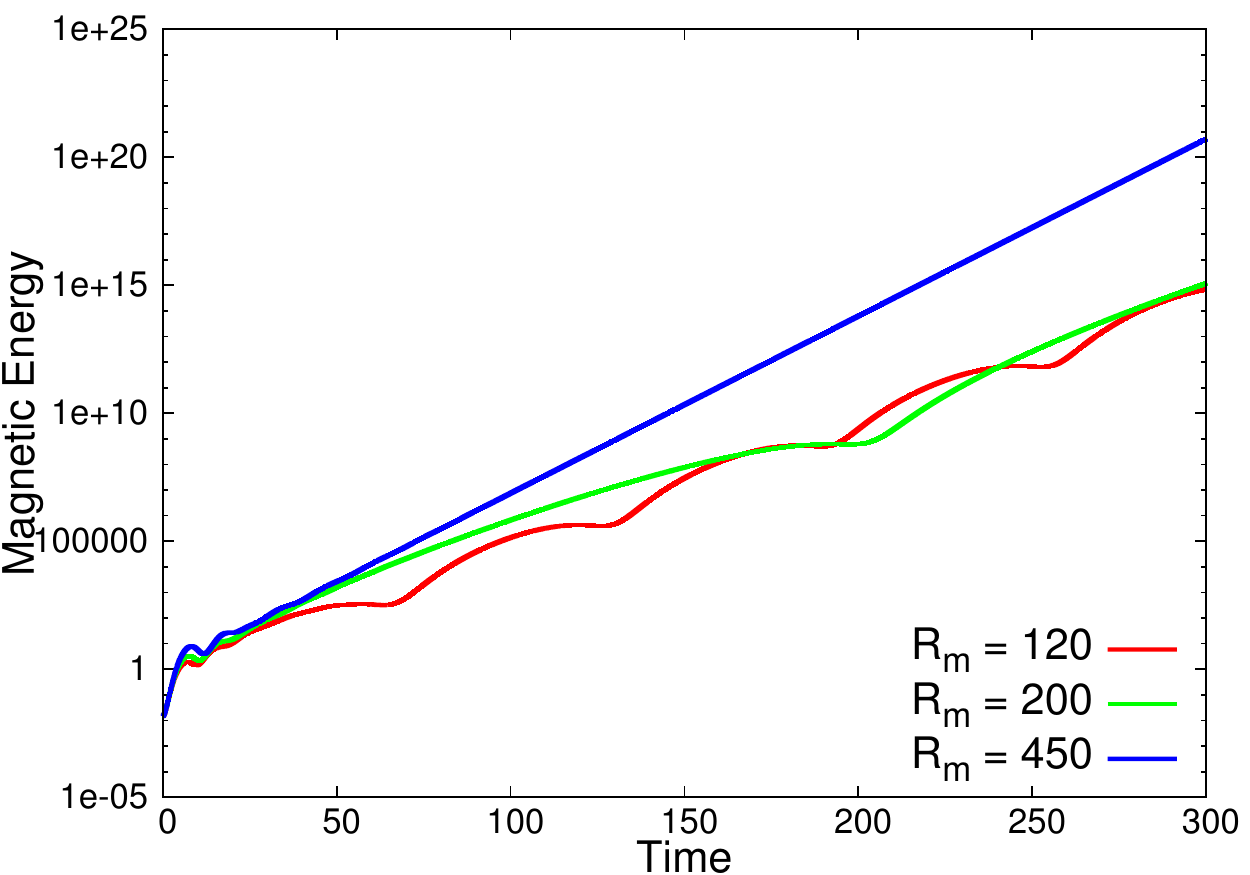}
\caption{Kinematic Dynamo effect reproduced using the identical parameter regime ($A = B = C = 1$, $k_f = 1$) by Galloway {\it et al} \cite{galloway:1986}. The grid resolution is $64^3$ which is close to the value $60^3$ that was taken by Galloway {\it et al} \cite{galloway:1986}. The growth rates of magnetic energy $\left(\sum\limits_{V} \frac{B^2(x,y,z)}{2}\right)$ of kinematic dynamo are found to increase as $Rm$ is increased. The oscillation frequency of the magnetic energy is also found to be similar as Galloway {\it et al} \cite{galloway:1986}}
\label{Frisch}
\end{figure}

Effect of magnetic resistivity ($\eta$) through the magnetic Reynold's number ($Rm$) has been widely studied in past \cite{galloway:1986, galloway:1986, galanti:1992} and in recent years \cite{bouya:2013}. First we reproduce the previous results by Galloway {\it et al} \cite{galloway:1986} using our code [Runs: KR1, KR2, KR3]. Similar to the previous study \cite{galloway:1986} we choose $U_0 = 1$, $A = B = C = 1$, $k_f = 1$. We time evolve only Eq. \ref{Bfield} for the initial time data mentioned above for magnetic Reynold's number $Rm = 120, 200, 450$ and obtain the identical growth of magnetic field as Galloway {\it et al} \cite{galloway:1986}. This result is shown in Fig \ref{Frisch}. We also reproduce the real and imaginary part of the eigenvalue obtained previously \cite{galloway:1986}. The critical value of onset of kinetic dynamo action is found to be $Rm = 27$.\\

\begin{figure}
\includegraphics[scale=0.65]{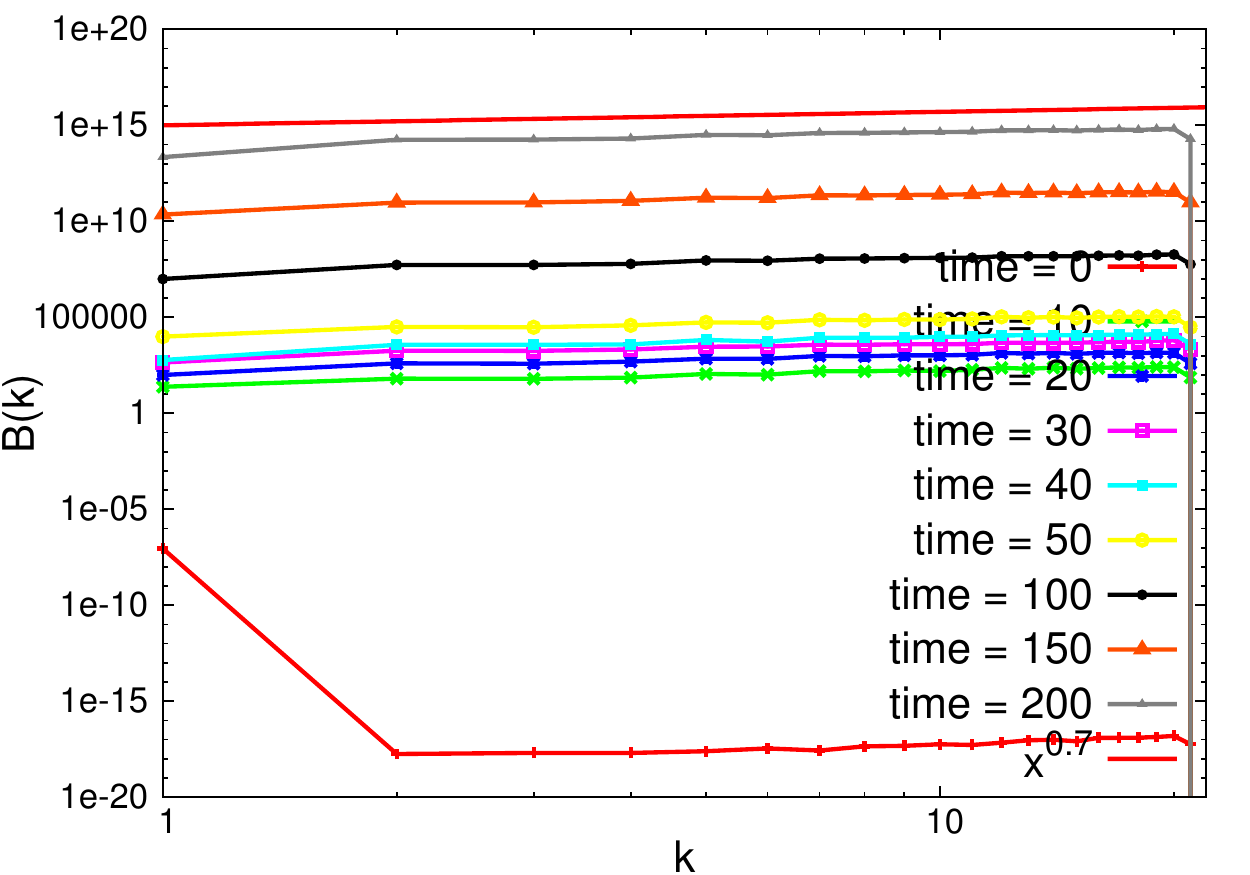}
\caption{Magnetic energy spectra at different time for ABC flow with the identical parameter described in Fig. (\ref{Frisch}) with $Rm = 450$. The energy contained in the large scales at late times shows that the short scales are equally important for a kinematic dynamo obtained from ABC flow.}
\label{Spectra_Frisch}
\end{figure}

We derive the energy spectra of the kinematic dynamo from ABC flow [Fig.(\ref{Spectra_Frisch})] and observe energy is not only contained in large scales rather, the energy contained in the intermediate scles are quite large. We observe a $k^{0.7}$ scaling of magnetic energy.

\subsubsection{Effect of Forcing Scale}
\label{subsubsec:forcing_scale}
\begin{figure}
\includegraphics[scale=0.65]{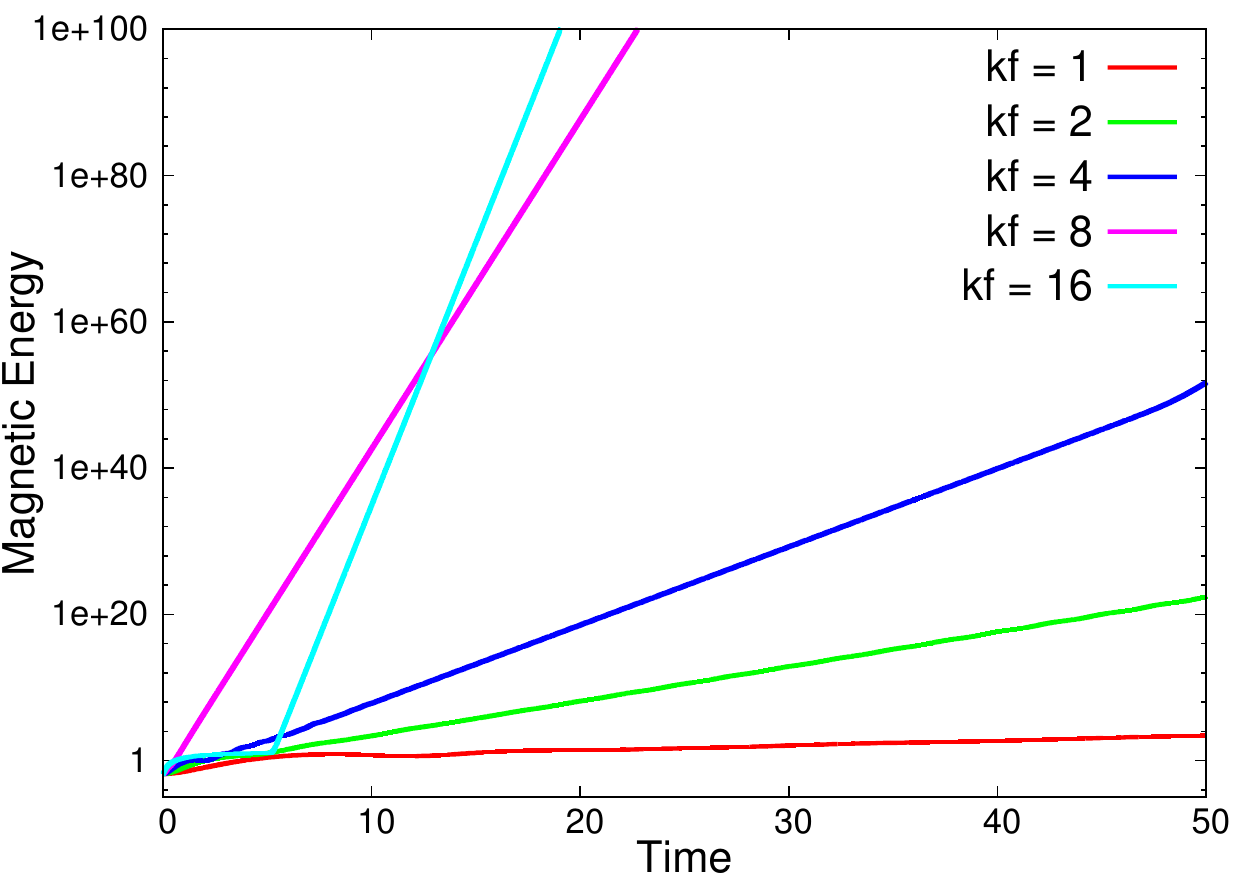}
\caption{Kinematic Dynamo effect for different driving frequency ($k_f$). The magnetic energy is normalised with the initial magnitude at time $t = 0$. The parameters chosen are $A = B = C = 1$ and $Re = Rm = 450$ with $U_0 = 1$. The initial growth rates are found to grow as $k_f$ is increased.}
\label{kin_kf_U0_1}
\end{figure}

\begin{figure}
\includegraphics[scale=0.65]{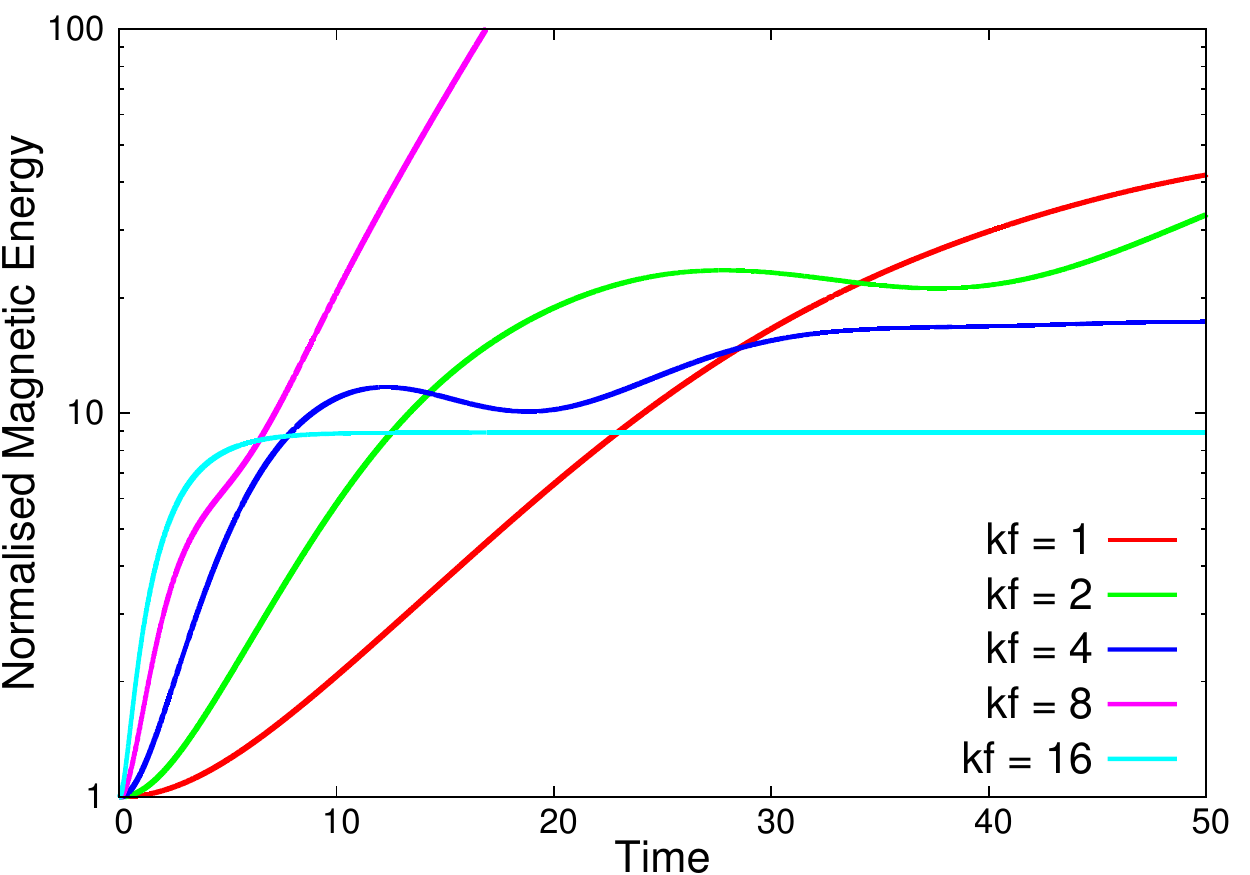}
\caption{Kinematic Dynamo effect for different driving frequency ($k_f$). The normalised magnetic energy is defined as $\left(\sum\limits_{V} \frac{B^2(x,y,z,t)}{2} - \sum\limits_{V} \frac{B^2(x,y,z,0)}{2}\right)$. The parameters chosen are $A = B = C = 1$ and $Re = Rm = 450$ with $U_0 = 0.1$. The initial growth rates are found to grow and saturate as the $k_f$ is increased.}
\label{kin_kf_initial}
\end{figure}

The effect of forcing scale on the growth rate of magnetic energy has been earlier studied by Galanti {\it et al} \cite{galanti:1992} for $k_f = 1$ to $10$ for $Rm$ values upto $Rm = 45$. For the kinematic dynamo case, Galloway and Frisch (\cite{galloway:1986}) have shown that, even though the critical value of $Rm$ for kinematic dynamo action for ABC flow is $Rm_c = 27$, the growth rate monotonically increases until $Rm = 350$. In the past work of Galanti {\it et al} \cite{galanti:1992} it was found that $k_f = 2$ has higher growth rate than $k_f =1$ for $Rm = 45$. Also changing $Rm$ from $12$ to $20$ did not affect the growth rate for different $k_f$ much. In our case, we keep the forcing length scale at $k_f = 1, 2, 4, 8, 16$ holding the $Rm = 450$ much above the critical value ($Rm_{crit} = 27$) of onset of kinetic dynamo for $k_f = 1$ (where imaginary part of the eigenvalue ($2 \gamma$) is undetectably small) and in the regime where the growth rate does not vary much with the further increment of $Rm$. [Runs: KF1, KF2, KF3, KF4, KF5] The growth rate of normalised magnetic energy, $\frac{B^2}{B^2_0}$, is found to increase as $k_f$ is increased [Fig.\ref{kin_kf_U0_1}, \ref{kin_kf_initial}]. However, the growth rate ($2 \gamma$) saturates as $k_f$ is increased for $U_0 = 0.1$[\ref{kin_kf_initial}. A similar saturation was also observed earlier though at $Rm = 12$ and $20$ \cite{galanti:1992}. 

The late time dynamics is found to be widely different for different driving frequencies ($k_f$) [Fig.\ref{kin_kf_U0_1},\ref{kin_kf_initial}]. For a kinetic dynamo problem similar transient behaviour ($k_f = 16$ and $8$ in Fig. \ref{kin_kf_initial}) starting from a typical initial condition has been addresses previously in detail \cite{bouya:2013}. It was found that when the fastest growing eigenmode is not excited, it takes some time for the fastest eigenmode to overcome the initially excited mode and hence the crossover happens at a later time. Even for a dynamic dynamo under external forcing, similar result was earlier obtained by Galanti {\it et al} \cite{galanti:1992} for $A = B = C = 1$, $k_f = 1$ and $Re = Rm = 12$.

\subsubsection{Effect of Alfven Speed}
\begin{figure}
\includegraphics[scale=0.65]{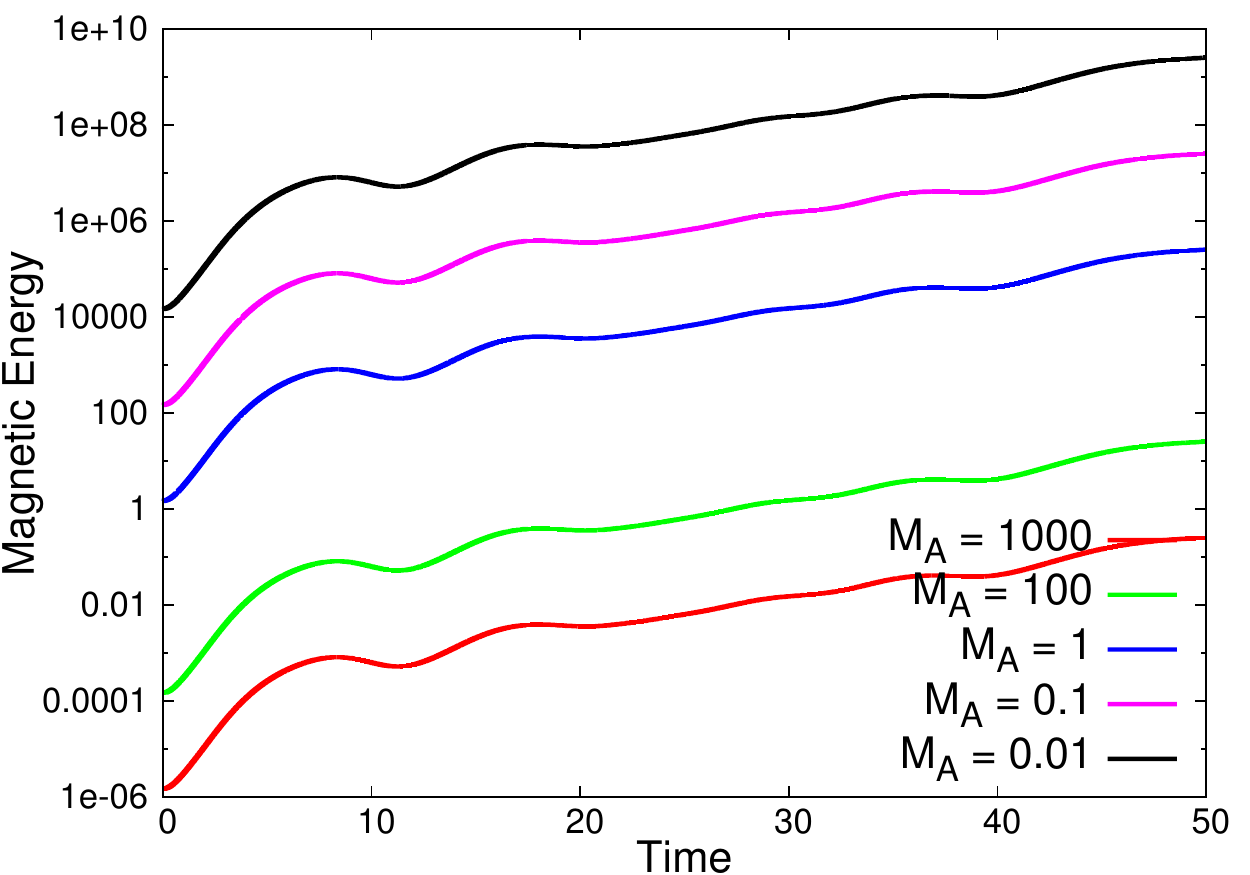}
\caption{Kinematic Dynamo effect for different driving frequency ($M_A$). The parameters chosen are $A = B = C = 1$ and $Re = Rm = 450$ with $U_0 = 1$. The growth rates are found to be identical for different $M_A$ values.}
\label{MA_kin_U0_1}
\end{figure}

\begin{figure}
\includegraphics[scale=0.65]{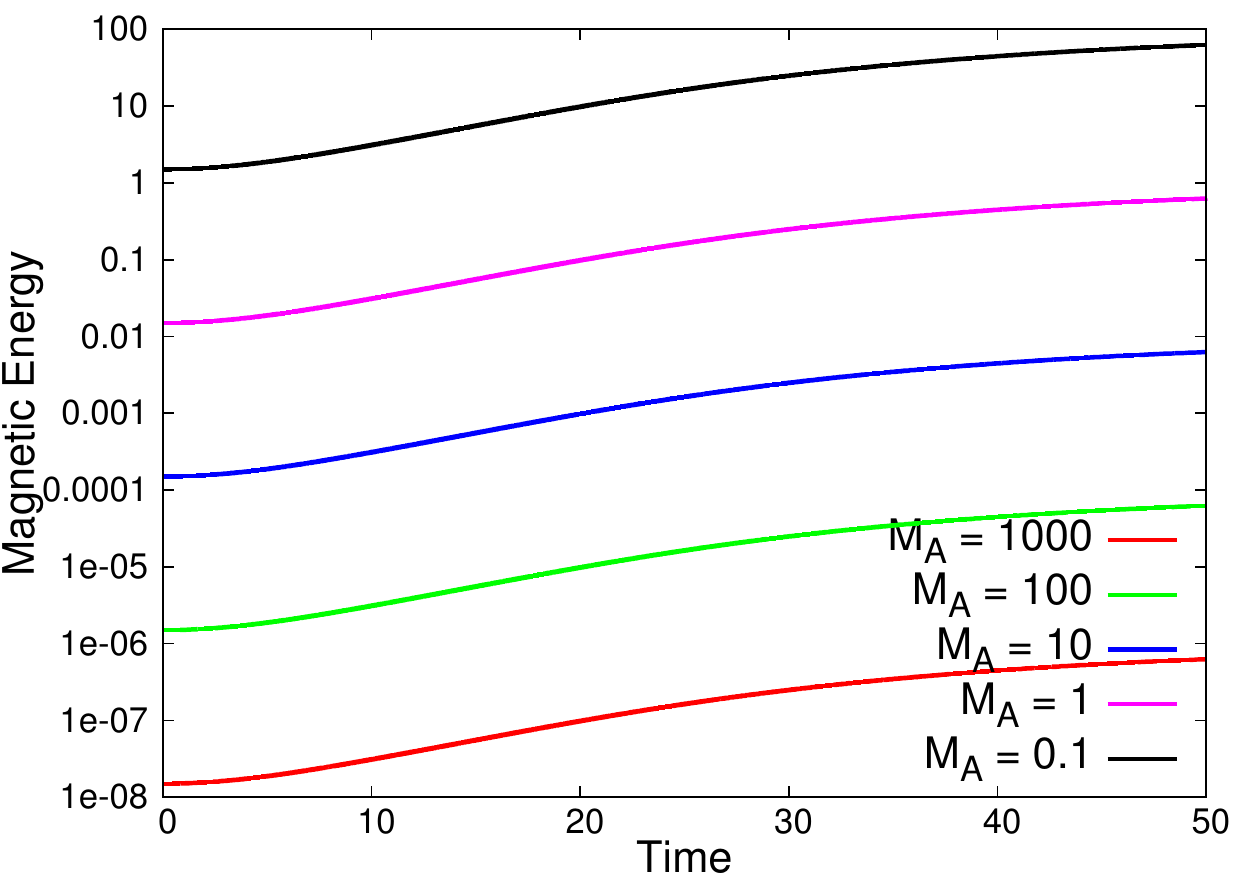}
\caption{Kinematic Dynamo effect for different driving frequency ($M_A$). The parameters chosen are $A = B = C = 1$ and $Re = Rm = 450$ with $U_0 = 0.1$. The growth rates are found to be identical for different $M_A$ values.}
\label{MA_kin}
\end{figure}

The Alfven speed is defined as $V_A = \frac{U_0}{M_A}$. If $M_A \textless 1$; $V_A \textless U_0$ and the plasma is called Sub-Alfvenic. Similarly if $M_A \textgreater 1$, the plasma is Super-Alfvenic. For kinetic dynamo problem, the growth rate of magnetic energy is found to be independent of the magnitude of $M_A$. [Runs: KF1, KM1, KR1, KM2, KM3] We check the growth rate for $M_A = 0.1, 1, 10, 100, 1000$ and for every case the growth rate of dynamo is found to be identical as shown in Fig. \ref{MA_kin_U0_1} and \ref{MA_kin} unlike the dynamic case discussed in the next subsection. 

\subsection{Dynamo with Back-reaction}
\begin{figure}
\includegraphics[scale=0.65]{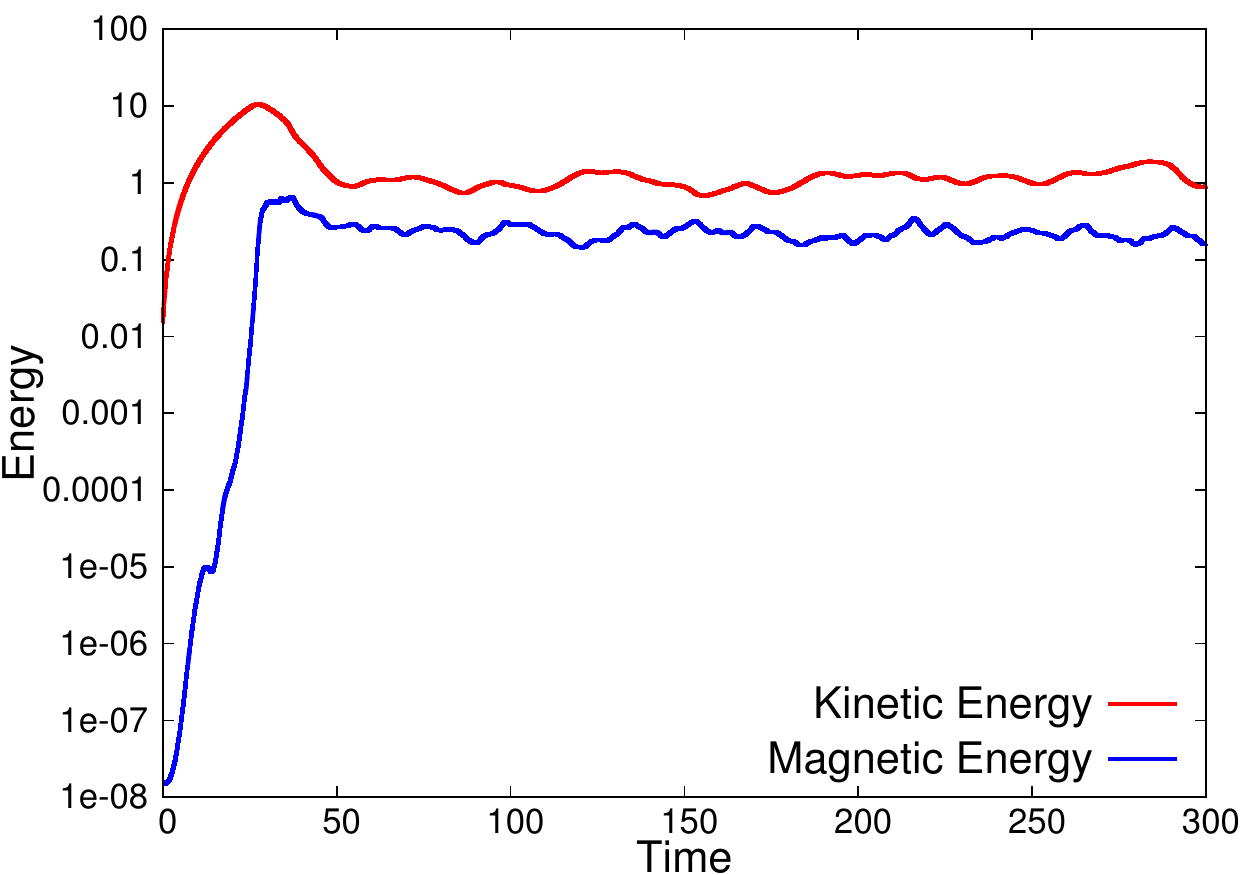}
\caption{Dynamic dynamo growth of kinetic $\left(\sum\limits_{V} \frac{U^2(x,y,z)}{2}\right)$ and magnetic $\left(\sum\limits_{V} \frac{B^2(x,y,z)}{2}\right)$ energy for ABC flow with $M_A = 1000$ and $k_f = 1$ for a long time with initial flow profile as ABC flow.}
\label{Long}
\end{figure}

A dynamic dynamo represents a situation where the magnetic energy grows exponentially for a plasma where the plasma itself evolves in time. Hence the velocity field is not externally imposed like a kinematic dynamo, rather it has a dynamical nature. The time evolution of the velocity field is generally governed by the Navier-Stokes equation including the magnetic feedback on the velocity field. In order to simulate such a scenario, we time evolve all the three equations, viz. Eq. \ref{density}, \ref{velocity}, \ref{Bfield}. A result for parameters $M_A = 1000$ and $k_f = 1$ for initial flow profile ABC is given in Fig. \ref{Long}.\\

We change the forcing scale, Alfven velocity and compressibility to observe the effect on the dynamics of the fields.\\

\begin{figure}
\includegraphics[scale=0.65]{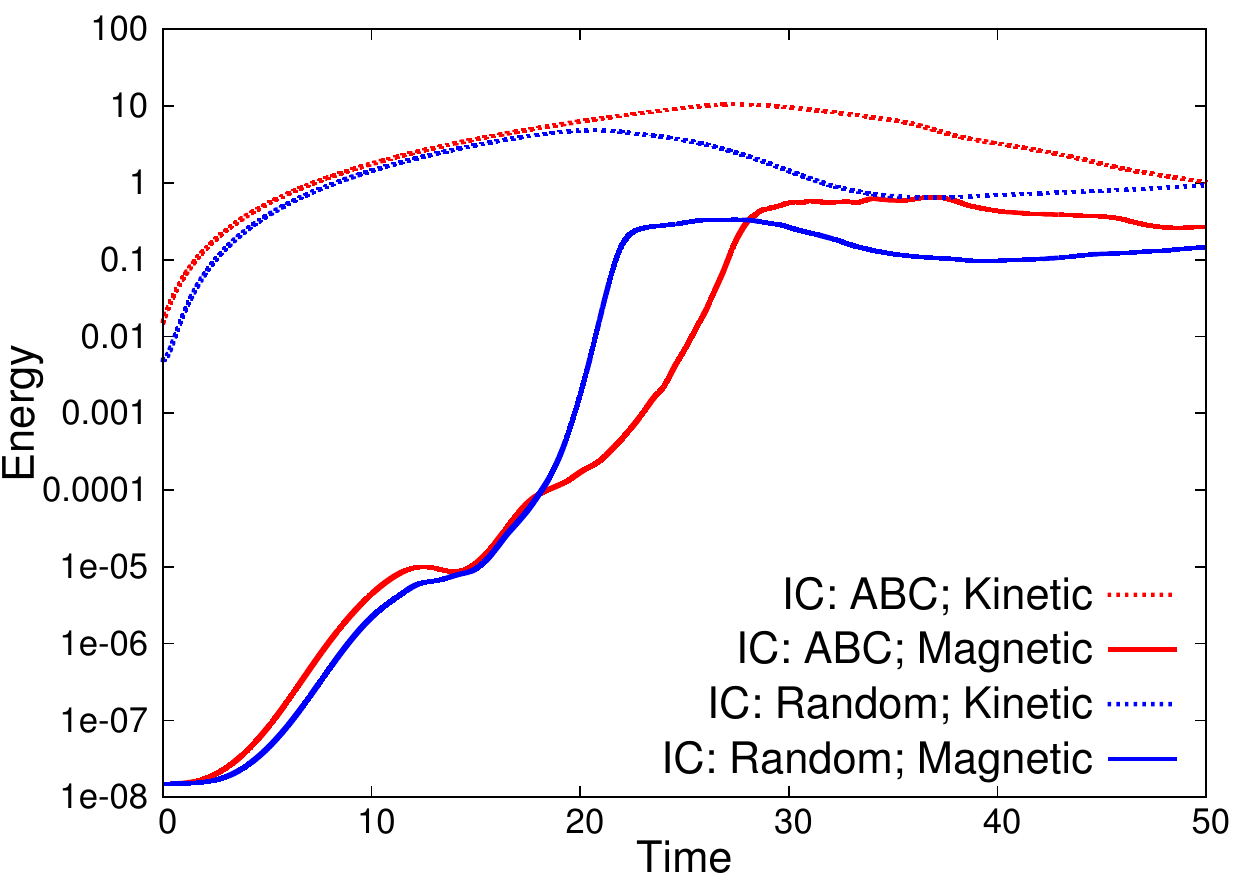}
\caption{Dynamic dynamo growth of kinetic and magnetic energy for two different initial conditions with $M_A = 1000$ and $k_f = 1$. The initial growth rate of magnetic energy for ABC initial flow profile is found to be identical with that of random field profile.}
\label{Rand}
\end{figure}

We turn on external forcing to the velocity field. We keep the nature of forcing as $f_x = A \sin(k_f z) + C \cos (k_f y)$, $f_y = B \sin(k_f x) + A \cos (k_f z)$, $f_z = C \sin(k_f y) + B \cos (k_f x)$. We keep $A = B = C = 0.1$ and $U_0 = 1$ throughout all the calculations and fix $Re = Rm = 450$. In case of an external forcing the initial memory is lost and hence the sensitivity to the initial condition is expected to be lost. We redo our numerical calculations for an initial random velocity field profile and find that the basic nature of dynamo effect does not get affected as shown in Fig. \ref{Rand}. The saturation regime for both the kinetic (sum over all velocity modes) and magnetic  (sum over all magnetic modes) energies remain the same though the two systems are evolved from different initial conditions. We perform the following runs (Table. \ref{parameter_run_for_dyn}) using our code to understand the externally forced ABC flow dynamo process.\\

\begin{table}[h!]
\centering
\begin{tabular}{ |c|c|c|c| }
 \hline
 Name & $k_f$ & $M_A$ & $M_s$ \\
 \hline
 FDF1 & $1$ & $1000$ & $0.1$ \\
 FDF2 & $2$ & $1000$ & $0.1$ \\ 
 FDF3 & $4$ & $1000$ & $0.1$ \\ 
 FDF4 & $8$ & $1000$ & $0.1$ \\ 
 FDF5 & $16$ & $1000$ & $0.1$ \\
 FDMA1 & $1$ & $100$ & $0.1$ \\
 FDMA2 & $1$ & $10$ & $0.1$ \\
 FDMA3 & $1$ & $1$ & $0.1$ \\
 FDMA4 & $1$ & $0.1$ & $0.1$ \\ 
 FDMS1 & $1$ & $100$ & $0.2$ \\ 
 FDMS2 & $1$ & $100$ & $0.3$ \\  
 FDMS3 & $1$ & $100$ & $0.4$ \\   
 FDMS4 & $1$ & $100$ & $0.5$ \\   
 \hline
\end{tabular}
\caption{Parameter details with which the simulation has been run for the externally forced dynamic dynamo problem.}
\label{parameter_run_for_dyn}
\end{table}

\begin{figure}
\includegraphics[scale=0.65]{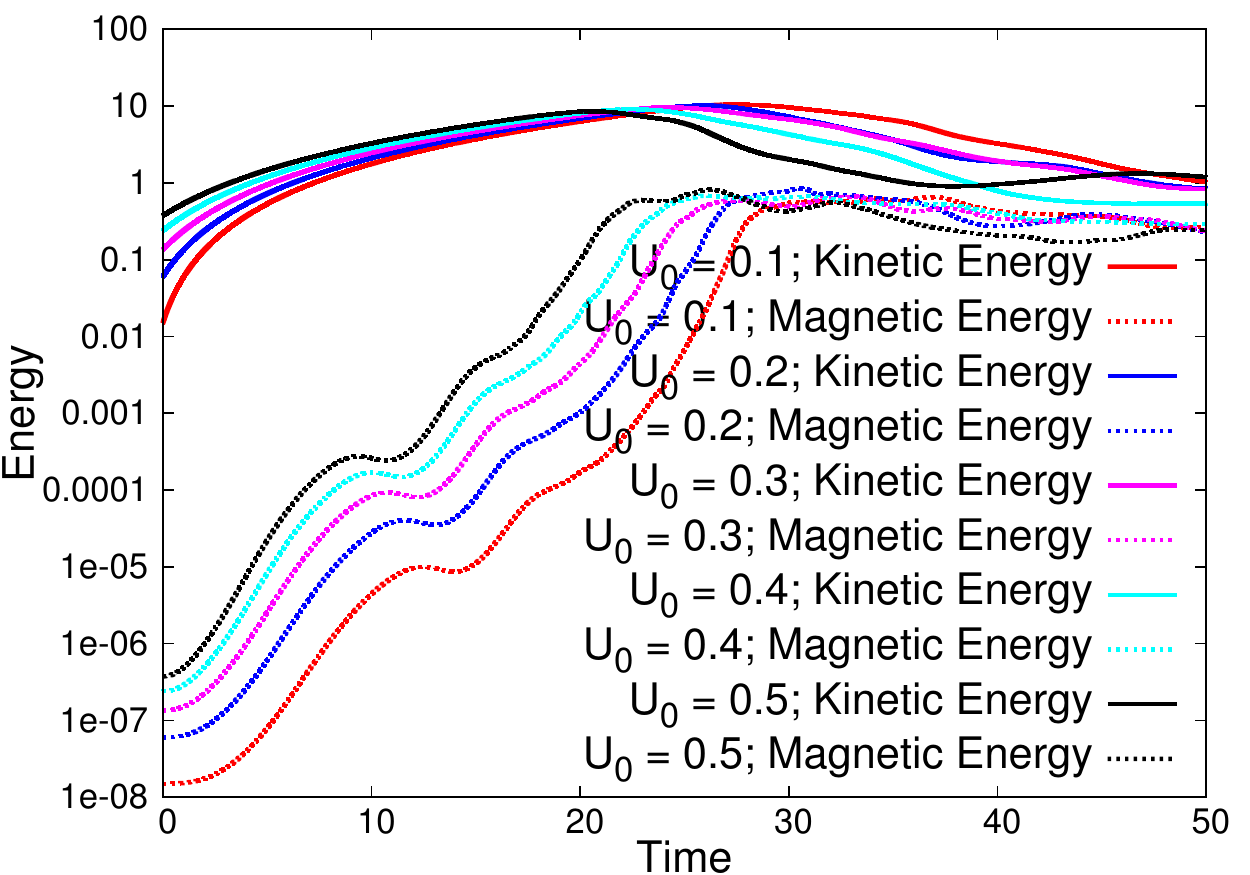}
\caption{Dynamic dynamo growth of kinetic and magnetic energy for five different initial velocities viz. $U_0 = 0.1, 0.2, 0.3, 0.4, 0.5$, with $M_A = 1000$ and $k_f = 1$. The initial growth rate of magnetic energy for ABC initial flow profile is found to be identical for all initial magnitude of velocities.}
\label{U_Comp_Forced}
\end{figure}

\begin{figure}
\includegraphics[scale=0.65]{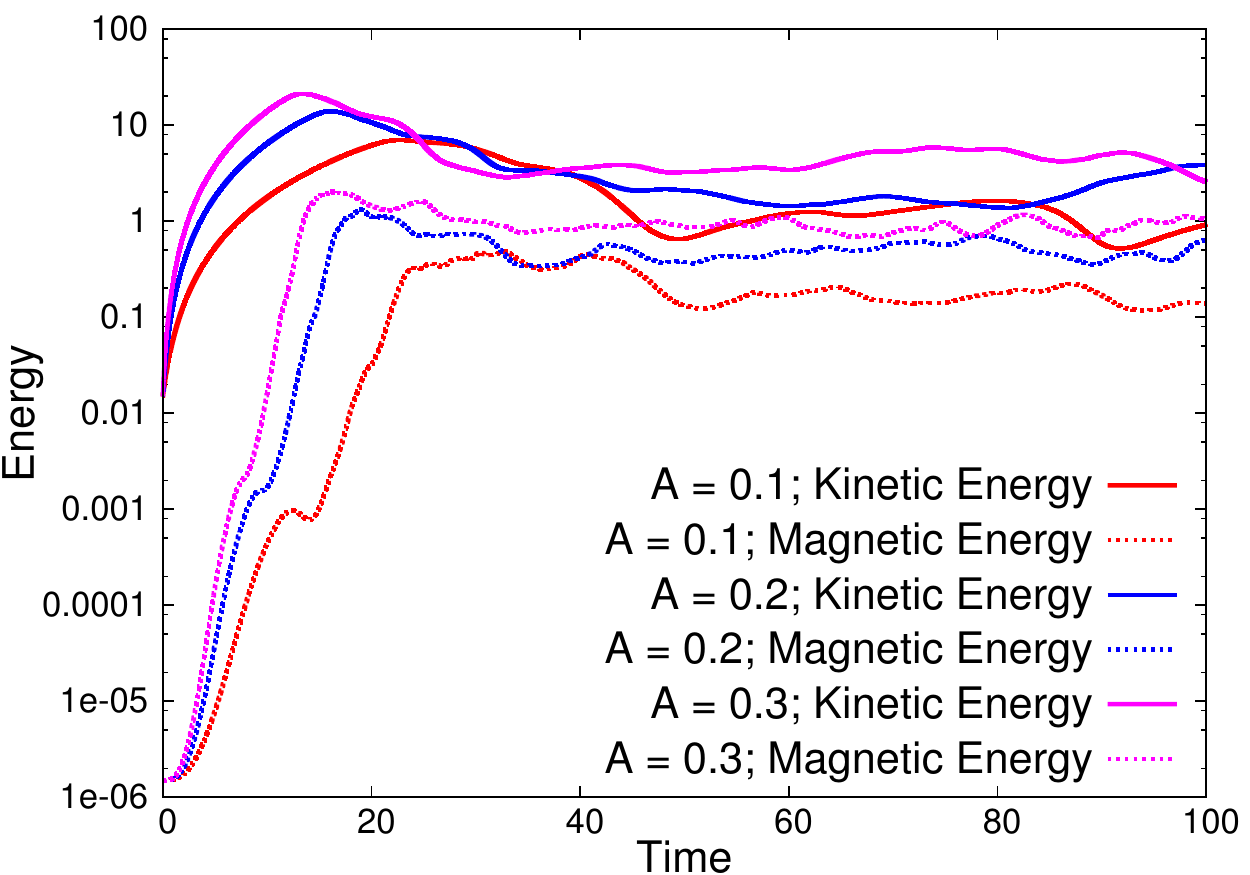}
\caption{Dynamic dynamo growth of kinetic and magnetic energy for three different forcing magnitudes $A = 0.1, 0.2, 0.3$ having $A = B = C$, with $M_A = 1000$ and $k_f = 1$. The initial growth rate of magnetic energy for ABC initial flow profile is found to be increase with the increament of magnitude of forcing.}
\label{ABC_Comp_Forced}
\end{figure}

Now we vary $U_0$ and the magnitude of $A, B, C$ keeping $A = B = C$ for all the cases. We run our simulation for $U_0 = 0.1, 0.2, 0.3, 0.4, 0.5$ keeping $A = B = C = 0.1$ and see the trend of dynamo action is identical for all values of $U_0$ [Fig. \ref{U_Comp_Forced}]. Next we vary the values of $A = 0.1, 0.1, 0.3$ keeping $A = B = C$ and $U_0 = 0.1$. We see faster growth of dynamo with higher values of forcing through the magnitudes of $A$, $B$ and $C$ [Fig.\ref{ABC_Comp_Forced}].

\subsubsection{Effect of Forcing scale}

\begin{figure}
\includegraphics[scale=0.65]{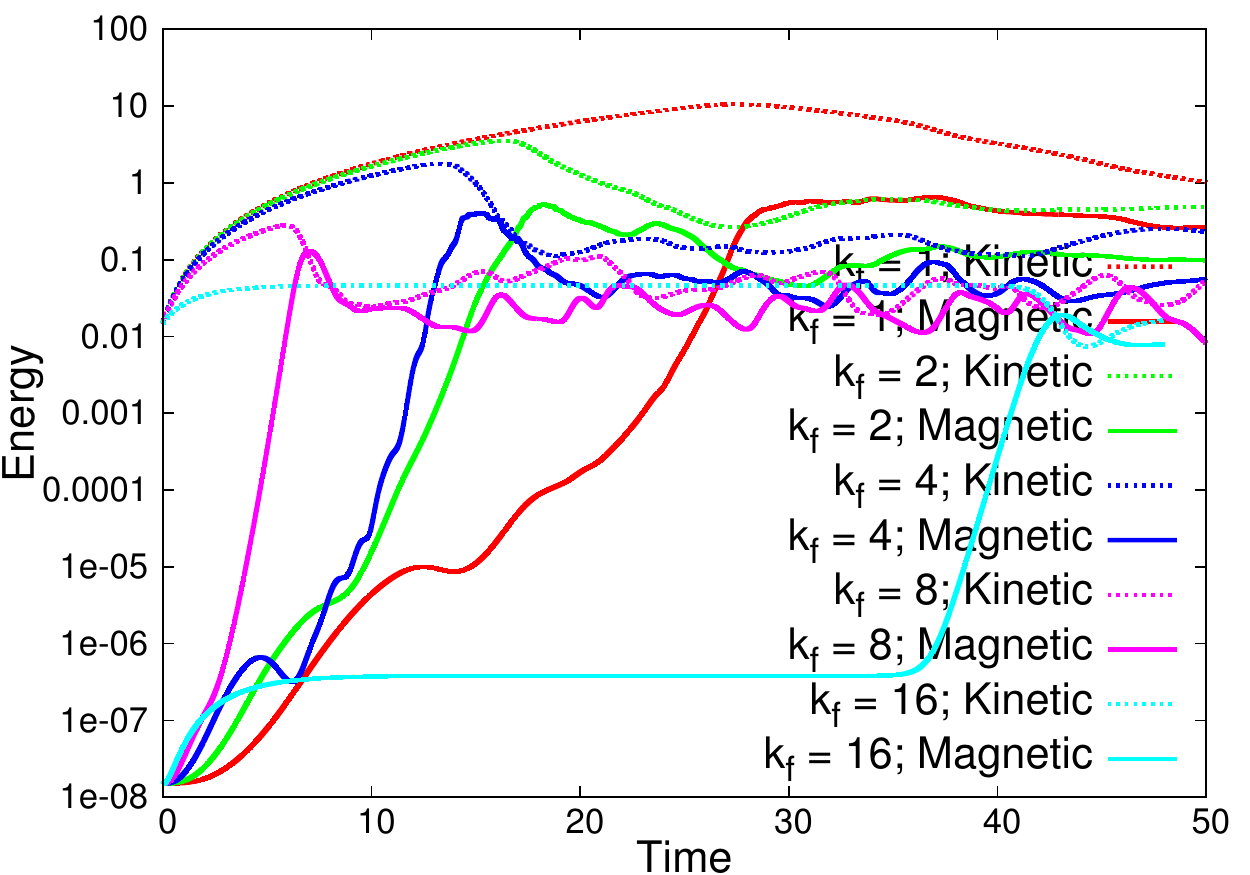}
\caption{Dynamic dynamo growth of kinetic and magnetic energy for $k_f = 1, 2, 4, 8, 16$ with $M_A = 1000$. The growth rate of magnetic energy is found to be steeper with the increase of $k_f$. The growth rate of kinetic energy due to external forcing decreases as $k_f$ is increased.}
\label{kf_F_MA_1000}
\end{figure}

A dynamic dynamo with external forcing has been studied earlier by Galanti {\it et al} \cite{galanti:1992} for incompressible plasma with $U_0 = 1$ $A = B = C = 0.1$, $Re$ \& $Rm$ upto $20$ (below $Rm_c = 27$), and $k_f = 1, 2, 4$. We change the length scale of forcing ($k_f$) on the velocity field keeping $U_0 = 0.1$, $A = B = C = 0.1$, $Re = Rm = 450$, $M_s = 0.1$ and $M_A = 1000$ as fixed parameters. [Runs: FDF1, FDF2, FDF3, FDF4, FDF5] From Fig. \ref{kf_F_MA_1000} we find, the growth rate of magnetic energy increases while that of kinetic energy decreases as $k_f$ is increased. The case $k_f = 16$ in Fig. \ref{kf_F_MA_1000} shows a delayed dynamo action. A possible explanation of this late time dynamo action is the excitation of a slow eigenmode to start with, which gets overpowered by the fastest eigenmode excited later. The identical phenomena we have seen in the kinematic dynamo section [Fig. \ref{kin_kf_initial}]. From Fig. \ref{kf_F_MA_1000} we also note that though externally forced, the saturation regime of both kinetic and magnetic eneries goes downwards as $k_f$ is increased. This is so, because the forcing scale also has a wave number term within it which helps to drain out energy through viscous dissipation, if a higher wavenumber ($k_f$) is excited. \\

\subsubsection{Effect of Alfven Speed}

We change the Alfven Mach number ($M_A$) of the plasma, keeping $U_0 = 0.1$ $A = B = C = 0.1$, $Re = Rm = 450$, $M_s = 0.1$ and $k_f = 1$. 

\begin{figure}
\includegraphics[scale=0.65]{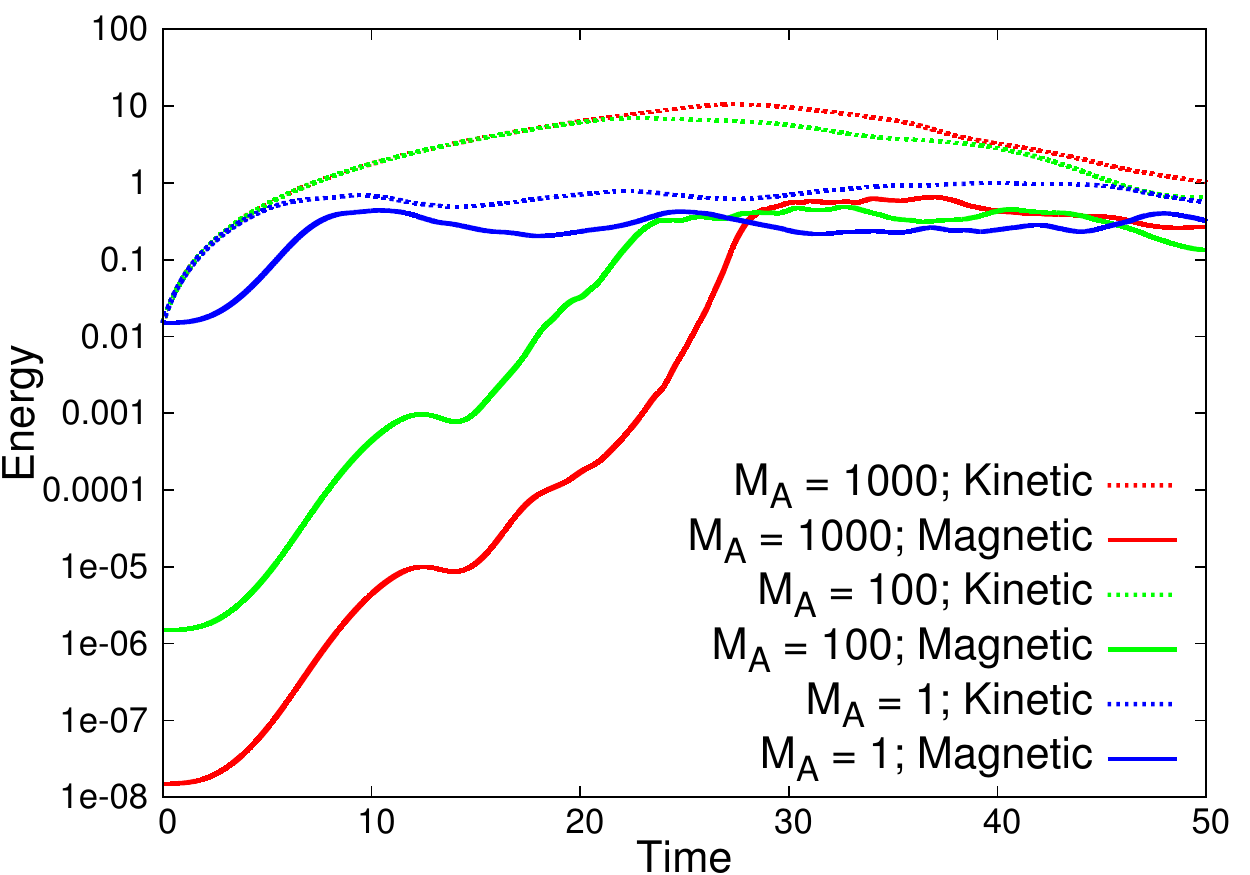}
\caption{Dynamic dynamo growth of kinetic and magnetic energy for $M_A = 1, 100, 1000$. The back-reaction of magnetic field on velocity field is found to affect the growth rate and dynamics of velocity field. This effect is captured in the time evolution of kinetic energy for different $M_A$.}
\label{feedback}
\end{figure}

We analyse the runs: FDF1, FDMA1, FDMA2, FDMA3. By choosing $M_A = 1, 10, 100$ and $1000$ we set the Alfven Velocity $V_A = \frac{U_0}{M_A} = 10^{-1}, 10^{-2}, 10^{-3}$ and $10^{-4}$ respectively. For $\rho_0 = 1$, $V_A = B_0$, the initial magnitude of the seed magnetic field profile. As we start from a lower value of $B_0$, the growth rate of the magnetic energy increases rapidly. This is quite similar to the kinematic dynamo action with a distinct difference. In kinematic dynamo there was no saturation of magnetic energy. On the other hand, in forced dynamic dynamo, there is a saturation value of the magnetic field. This saturation is believed to be due to the backreaction of the magnetic field on the velocity field through the Lorentz force term. The strong magnetic field generated through the dynamo process, starts affecting the topology of the velocity field in turn affecting its dynamics. Thus the modified velocity field no longer remains a ABC flow and finally the dynamo saturates. The effect of such magnetic feedback on the velocity field is shown in Fig \ref{feedback} for $M_A = 1,100, 1000$. \\

We do the following observations from Fig \ref{feedback}. 

$\bullet$ We notice that, for both the case $M_A = 100$ and $1000$ there exists three distinct slopes. At the beginning, the magnetic energy starts exponentialy increasing with time. Once it gets amplified by around four orders of magnitude, the exponent of increament suddenly falls down for both the cases $M_A = 100$ and $1000$. After that, the magnetic energy again starts increasing with higher exponent. 

$\bullet$ It is also note-worthy that, the initial growth rate of the magnetic energy for $M_A = 100$ and $1000$ are identical though they differ later on. Thus we understand it as similar to Kinematic dynamo (\ref{MA_kin}) where the backreaction is negligible. However, at later time because of the difference in the strength of the backreaction, the slopes of increament of the magnetic energy in logarithmic scale differs. 

$\bullet$ We also find that when the growth of the dynamo is several orders of magnitude (for higher values of $M_A$) the kinetic energy also grows faster though ultimately both kinetic and magnetic energies saturate at the same value. 

$\bullet$ We see that, independent of the strength of the seed magnetic field, the saturation regime of the kinetic and magnetic energies are the same. 

Thus we conclude from the above observations that, if the velocity field is ABC forced, whatever be the seed magnetic field, the dynamo effect becomes possible in super Alfvenic systems and the dynamo action is quite strong leading the final magnetic energy comparable to the kinetic energy.\\

\subsubsection{Energy Spectra}

Now we analyse the kinetic and magnetic energy spectra of the dynamic dynamo action at different times for $U_0 = 0.1$, $A = B = C = 0.1$, $k_f = 1$, $M_A = 1000$, $Re = Rm = 450$. Initially the energy content was limited to the fundamental mode only. But in course of time the kinetic energy shows a $k^{-5/3}$ spectra while the magnetic energy shows a $k^{0.7}$ spectra identical to the kinematic dynamo phenomena. However it is worth notable that the growth of magnetic energy in intermediate scales is much slower than the kinematic dynamo as can be found in Fig.\ref{Spectra_Frisch}\\

\begin{figure}
\includegraphics[scale=0.65]{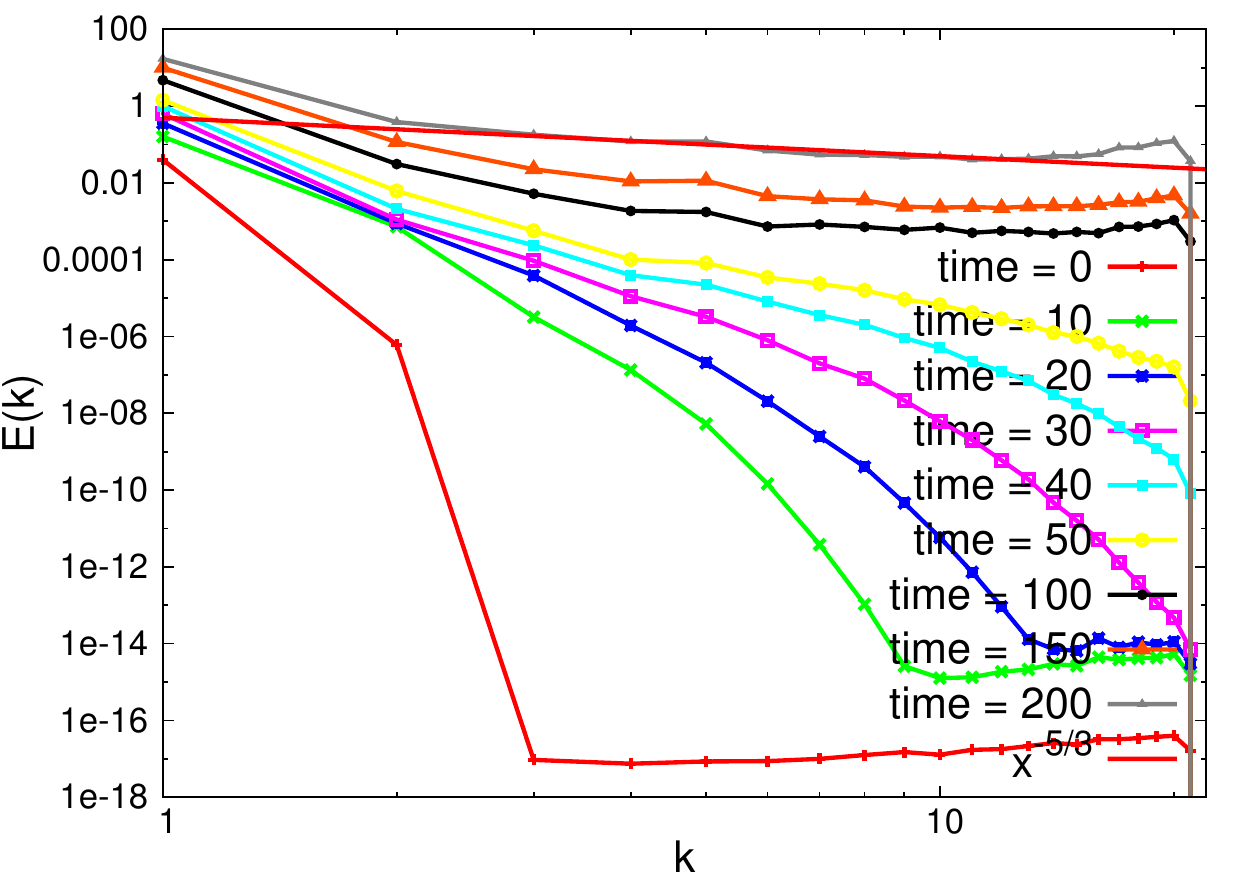}
\caption{Kinetic energy spectra for dynamic dynamo with $U_0 = 0.1$, $A = B = C = 0.1$, $k_f = 1$, $M_A = 1000$, $Re = Rm = 450$.}
\label{E_Spectra}
\end{figure}
\begin{figure}
\includegraphics[scale=0.65]{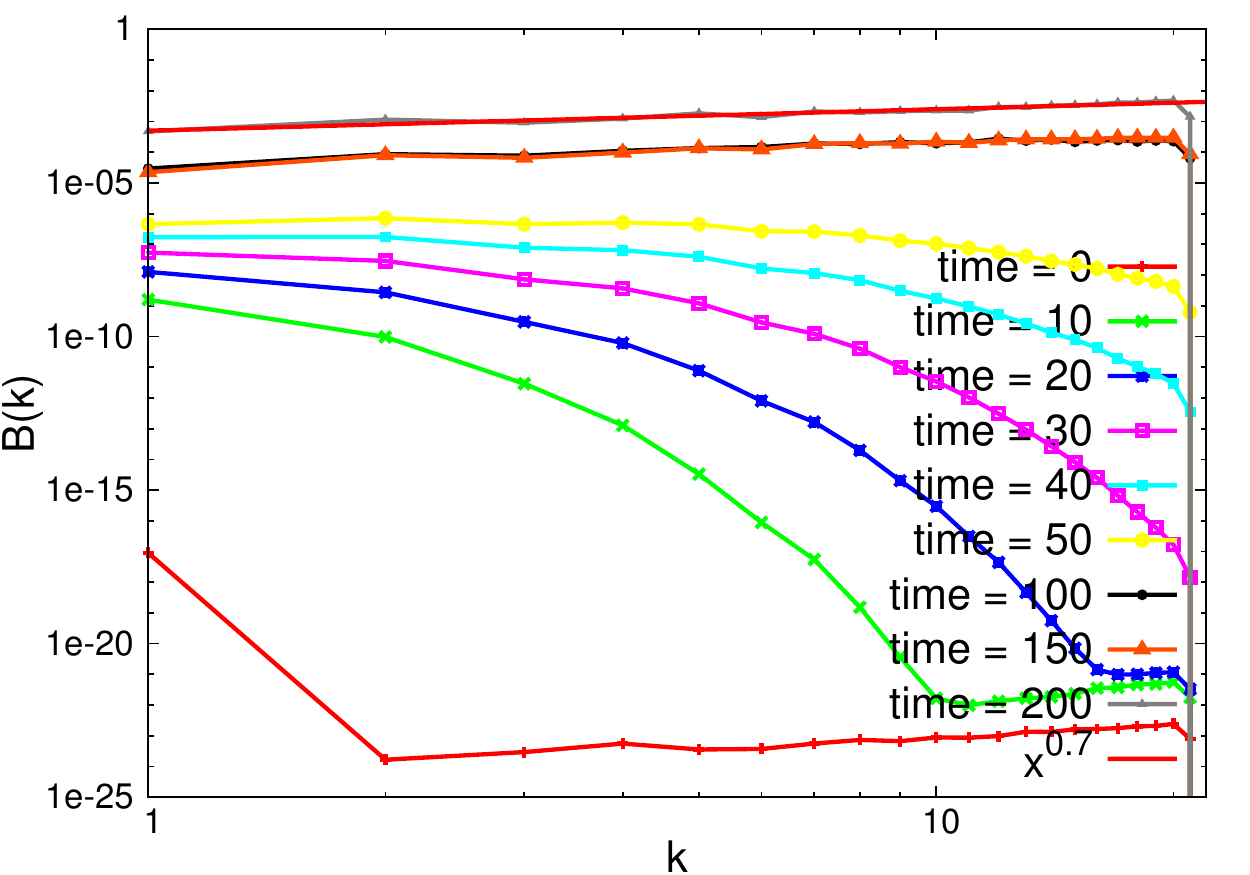}
\caption{Kinetic energy spectra for dynamic dynamo with $U_0 = 0.1$, $A = B = C = 0.1$, $k_f = 1$, $M_A = 1000$, $Re = Rm = 450$.}
\label{B_Spectra}
\end{figure}



 


\section{Summary and Future Works}
In this work we have analysed several phenomena of a magnetohydrodynamic plasma under ABC flow.\\

$\bullet$ First we study the kinematic dynamo effects where the velocity field is the ABC flow - a known solution of Euler equation. At different wave-numbers of the flow, we see that the growth rate of the magnetic energy in the kinematic dynamo case increases as $k_f$ is increased.\\
 
$\bullet$ In case of an ABC forced velocity field for the dynamic dynamo problem seems to show similar variation with $k_f$, though now the dynamo action becomes very prominant. The magnetic energy grows upto the order of kinetic energy when we remain in the super Alfvenic regime.\\

$\bullet$ The magnetic energy is found to be contained primarily in the intermediate scales in wave-number.\\

The compressibility however has not been found to affect the results for the weakly compressible cases. The effect of variation of initial density ($\rho_0$) on the dynamo effect can be an interesting piece of study and will be explored elsewhere.\\


\section{Acknowledgement}
R.M. acknowledges several insightful discussions with Akanksha Gupta, Vikrant Saxena and Abhijit Sen at Institute for Plasma Research, India. The development as well as benchmarking of MHD3D has been done at Udbhav and Uday clusters at IPR.



\bibliography{biblio}

\end{document}